\documentstyle[11pt]{article}
\normalbaselines
\begin{document}

\begin{center}
{\Large \bf
Multidimensional Phase Space and Sunset Diagrams} \\ [5mm]

A. Bashir{\footnote{\it Permanent address: University of Michoacan, Morelia,
MEXICO~~(E-mail: adnan@itzel.ifm.umich.mx)}},
R. Delbourgo{\footnote {\it E-mail: Bob.Delbourgo@utas.edu.au}}
and M.L. Roberts{\footnote {\it E-mail: Martin.Roberts@utas.edu.au}}\\
[5mm]
{\it School of Mathematics and Physics, University of Tasmania, Hobart, 
AUSTRALIA 7001}
\end{center}

\begin{abstract}
We derive expressions for the phase-space of a particle of momentum $p$ 
decaying into $N$ particles, that are valid for any number of dimensions. 
These are the imaginary parts of so-called `sunset' diagrams, which we also 
obtain. The results are given as a series of hypergeometric functions, which
terminate for odd dimensions and are also well-suited for deriving the 
threshold behaviour.
\end{abstract}

\section*{I. Introduction}
With so much attention focussed on the properties of branes embedded in 
higher dimensions, it is of interest to examine the way in which the phase 
space changes as the space-time is enlarged, since this is what primarily 
determines the statistical properties of multiparticle systems; only when 
amplitudes are accentuated or suppressed along certain directions in space 
is there a pronounced effect on the phase space. This paper is devoted to 
the topic of multidimensional phase space. We assume that extra dimensions 
are spatial---though one may envisage that time-like coordinates can be 
handled by Euclidean continuation---and that all particles propagate into 
the bulk. Not only that, we suppose that the space-time is flat so that we 
can neglect topological aspects. If conditions arise that constrain the 
particles to some subspace, then one may obtain the reduced phase space by 
altering the number $D-1$ of spatial dimensions rather trivially; or else if
topological effects become important we may be able to sum over discrete
modes along flat directions. Thus in almost every respect we are simply 
investigating the dimensional continuation of standard 4-D expressions for 
the decay of a system carrying momentum $p$ into {\em any number} $N$ of 
particles.

There has been much early work in this area \cite{HAK}, but it is mostly in 
the context of four dimensions (or lower) and has often been confined to few 
particles (like two, three or four). The most pertinent recent results in 
this connection are those of Groote {\em et al} \cite{GKP} and of Davydychev
and Smirnov \cite {DS} because they are the most general. Our own results 
apply to any value of $N$ and $D$ and
are also singularly well-suited for deriving threshold expansions 
as $\sqrt{p^2}\rightarrow (m_1+m_2+\cdots + m_N)$. Because phase space 
$\rho$ is nothing but the imaginary part of the `sunset diagram' for 
$p\rightarrow p_1+p_2+\cdots p_N$, our procedure will be to 
start with this sort of diagram and then obtain $\rho_N$ from its 
discontinuity in $p^2$ as we cross the threshold. (Pseudothreshold 
singularities of these diagrams exist too but are not relevant for 
physical phase space.)

In Section II, we give the most convenient Feynman parametric description of
the general diagram for any dimension $D=2\ell$ and show how one may readily
derive the leading threshold behaviour by expanding about the minimum of the
combined denominator; although one may in principle derive the next to 
leading behaviour by this method, and so on, we do not pursue this 
parametric approach any further, as there is a much better way of handling 
the problem, which is described in the following sections. First, in
section III, we derive some special cases of the results (small $N$) by 
making use of phase space recurrence methods \cite{HAK} and next show how 
they may more easily be found by Fourier transformation of the propagator 
products in section IV. This is our cue for tackling the most general 
situation, but because the form of our expressions makes it tricky to take 
the massless limit of any individual particle we specifically suppose that 
$M$ of the particles are massless and $N-M$ are massive. Our results, stated
in section V, are given as a series of hypergeometric terms having the form
(herafter $F$ stands for the usual $_2F_1$ function)
$$\rho_N \propto \sum_{j=M(3/2-\ell)+2N(\ell-1)-\ell+1/2}d_j(p^2-\sigma^2)^j
 \,\,F(a,b;c;1-p^2/\sigma^2),$$
where $\sigma$ is the sum of the masses. An agreeable property of this 
expansion is that it terminates for odd $D$ and is tailored for deriving
the threshold behaviour.

\section*{II. Feynman parametric form}
In 2$\ell$ dimensions, sunset diagrams with $N$ internal lines, produce 
integrals of the type\footnote{The bar notation means: division by $2\pi$
when integrating over each momentum component and a factor of 2$\pi$ for 
each delta function.},
\begin{equation}
I_N(p;\{\nu\})=i\left(\prod_{i=1}^N\int\frac{i\Gamma(\nu_i)}
{(p_i^2-m_i^2)^{\nu_i}}\bar{d}^{2\ell}\!p_i\right)
\bar{\delta}^{2\ell}(p-\sum_i p_i). 
\end{equation}
We have included a Gamma function in each numerator since the propagator 
powers usually arise via momentum or mass derivatives of the case $\nu = 1$,
and it simplifies the subsequent expressions. By attaching a Feynman 
parameter $\alpha$ to each internal line, one may combine denominators in 
the standard fashion and integrate over the internal momenta to
establish\footnote{This involves some finicky rescaling of the Feynman 
parameters. See reference \cite{BDU} for the case $N=3$.} that
\begin{equation}
 I_N(p;\{\nu\}) = \frac{(-1)^N}{(-4\pi)^{(N-1)\ell}}\left(\prod_{i=1}^N
           \!\int_0^1\!\frac{d\alpha_i}{\alpha_i^{\nu_i-\ell+1}}\right)
           \frac{\Gamma(\sum_i\nu_i+\ell-N\ell)\delta(1-\sum_i\alpha_i)}
                {[p^2\!-\!\sum_i(m_i^2/\alpha_i)]^{\sum_i\nu_i+\ell-N\ell}},
\end{equation}
but it is easier to prove the result by induction in fact. Suppose that (2)
is true for $N$; the case $N+1$ represents a further convolution:
$$I_{N+1}(p;\{\nu\})=i\int \bar{d}^{2\ell}\!k \,I_N(k;\{\nu\})
                     \Gamma(\nu_{N+1})/[(k+p)^2-m_{N+1}^2]^{\nu_{N+1}} $$
so introduce an extra Feynman parameter $\beta$. Combining the new 
denominator with (2) and integrating over intermediate loop momentum $k$,
one remains with
$$I_{N+1}(p)\!=\!\frac{(-1)^{N+1}}{(-4\pi)^{N\ell}}\!\left(\!\prod_{i=1}^N
 \int_0^1\!\!\!\frac{d\alpha_i}{\alpha_i^{\nu_i-\ell+1}}\!\right)
 \!\!\int_0^1\!\!\!d\beta \frac{\Gamma(\sum_{i=1}^{N+1}\nu_i\!-\!N\ell)
 \delta(1\!-\!\sum_i\alpha_i)(1\!-\!\beta)^{\nu_{N+1}-1}
 \beta^{\ell-\nu_{N+1}-1}}{[p^2(1\!-\!\beta)\!-\!\sum_i 
 (m_i^2/\alpha_i)\!-\!m_{N+1}^2(1\!-\!\beta)/
 \beta]^{\sum_i\nu_i-N\ell+\nu_{N+1}}}.$$
Now all one need do is rescale $\alpha_i=\beta_i/(1-\beta)$ for $i=1$ to $N$
and call $\beta\equiv\beta_{N+1}$. A final relabelling of $\beta\rightarrow
\alpha$ reproduces (2) for $N\rightarrow N+1$. Since we know the result (2)
is correct for $N=2$ ---it is very familiar to graphologists as the simplest
self-energy calculation---we have thereby proved the result inductively for
any $N$.

The singularities of $I_N(p)$ in $p^2$ will arise when $p^2$ equals the
combination $M^2(\alpha)\!\equiv\!\sum_{i=1}^N(m_i^2/\alpha_i)$, so let us
examine the behaviour of $M^2(\alpha)$ in the region of integration 
$0\leq\alpha_i\leq 1$, subject to the condition $\sum_i\alpha_i=1$. For
definiteness, we shall assume for the rest of this section that all masses 
are nonzero. By introducing a Lagrangian multiplier for the last constraint,
it is easy to see that $M^2(\alpha)$ is minimised in the physical region 
when the Feynman parameters equal $\alpha_{i0}\!\equiv\!m_i/
\sum_{j=1}^N m_j\equiv\!m_i/\sigma$, whereupon $M^2(\alpha_0)=(\sum_{j=1}^N 
m_j)^2=\sigma^2$. Other values of $\alpha_i$ are possible (by reversing one
or other of the signs of $\alpha_{i0}$) but they correspond to 
pseudothresholds and lie outside the integration region. It is
then clear that the magnitude of $I_N(p)$ will be dominated by $\alpha_i$ 
values in the vicinity of $\alpha_{i0}$, so it is sensible to expand about 
these minima if we want to determine the leading behaviour near threshold. 

This procedure can be developed systematically (see the Appendix for the 
example $N=2$ with $\nu_1=\nu_2=1$) but as there is an alternative and 
preferable way of tackling the problem, which we defer to the following 
sections, here we shall simply extract the leading threshold behaviour, 
because this can be done rather quickly and painlessly. Begin with (2) and 
note that
$$M^2(\alpha)=\sigma^2+\sum_i(\alpha_i-\alpha_{i0})^2\sigma^3/m_i-
           \sum_i(\alpha_i-\alpha_{i0})^3\sigma^4/m_i^2+\cdots$$
One sees that for $1-p^2/\sigma^2\equiv\Delta^2$ small the integral is 
dominated by values of $\alpha$ near the minimum $\alpha_0$. Therefore write
$\alpha_i = \alpha_{i0}+\Delta\xi_i$, so that the denominator of (2) reads
$$-\Delta^2\sigma^2[1+\sigma\sum_i\xi_i^2/m_i+
   \Delta\sigma^2\sum_i\xi_i^3/m_i^2 +O(\Delta^2)].$$
Since the integrals in $\alpha$ can be made to run between 0 and $\infty$,
because of the delta-function constraint, the integral over $\xi_i$ runs
from $-m_i/\sigma\Delta$ to $\infty$. So, in the limit as $\Delta\rightarrow
0$, we can take $\xi_i$ to run from $-\infty$ to $\infty$ (the correction
to the integral is exponentially small as $\exp(-1/\Delta^2)$) and expand
the product of the $\alpha_i$ about the minimum. Assuming all masses are
non-zero (see later sections for a relaxation of this condition) we are 
thereby led to the leading behaviour,
\begin{equation}
I_N(p,\{\nu\})= c_{N\ell}\Delta^{(N-1)(2\ell+1)-2\sum_i\nu_i}(\prod_i 
 m_i^{\ell-\nu_i-1})/(4\pi)^{\ell(N-1)}\sigma^{(2-N)\ell-N+\sum_i\nu_i},
\end{equation}
where
$$c_{N\ell}\simeq\left(\prod_i\int\!\!d\xi_i\right)(-1)^{N-\sum_i\nu_i}
 \frac{\delta(\sum_i\xi_i)\Gamma(\sum_i\nu_i-(N-1)\ell)}
 {[1+\sigma\sum_i\xi_i^2/m_i]^{\sum_i\nu_i-(N-1)\ell}}.$$
If one now represents the delta function and the denominator by integrals,
the coefficient $c_{N\ell}$ can be explicitly evaluated as follows:
\begin{eqnarray*}
c_{N\ell}&=&\frac{1}{2\pi}\left(\prod_i\int d\xi_i\right)
            \int_{-\infty}^\infty dk\int_0^\infty d\alpha\,\,
       \alpha^{\sum_i\nu_i-1-(N-1)\ell}\,{\rm e}^{-\alpha-k^2/4\alpha}\\
         & & \qquad\times (-1)^{N-\sum_i\nu_i}\,\,\prod_i\exp
        \left[-\frac{\alpha\sigma}{m_i}
             (\xi_i-\frac{ikm_i}{2\alpha\sigma})^2 \right]\\
 &=& (-1)^{N-\sum_i\nu_i}\,\varpi^{1/2}\pi^{(N-1)/2}\sigma^{-N/2}
      \Gamma(\sum_i\nu_i-(\ell+1/2)((N-1)); \qquad \varpi\equiv \prod_i m_i.
\end{eqnarray*}
So, all told, the leading threshold behaviour is dominated by
\begin{equation}
I_N(p,\{\nu\})\!=\!(-1)^{N-\sum_i\!\!\nu_i}
                \frac{\pi^{(N-1)/2}}{(4\pi)^{(N-1)\ell}}
 \frac{\Gamma(\sum_i\!\!\nu_i-(N\!-\!1)(l+1/2))\prod_im_i^{\ell-\nu_i-1/2}}
 {\sigma^{(2-N)\ell-N/2+\sum_i\nu_i}}\Delta^{(N-1)(2\ell+1)-2\sum_i\nu_i}.
\end{equation}
The case of greatest interest is $\nu_i=1$, all $i$, when (4) reduces to 
\begin{equation}
I_N(p)=\frac{\pi^{(N-1)/2}}{(4\pi)^{(N-1)\ell}}
 \frac{\Gamma((N-1)/2-\ell(N-1))}{\varpi^{3/2-\ell}\sigma^{N/2+\ell(2-N)}}
(1-p^2/\sigma^2)^{\ell(N-1)-(N+1)/2}.
\end{equation}
We may then continue this expression above threshold ($p^2\geq\sigma^2$), in
order to obtain the behaviour of the $N$-body phase space as the
discontinuity:
\begin{equation}
\rho_N(p)=2\Im I_N(p)\simeq\frac{\pi^{(N+1)/2}}{(4\pi)^{(N-1)\ell}}
       \frac{\varpi^{\ell-3/2}}{\sigma^{N/2+\ell(N-1)}}
       \frac{(p^2/\sigma^2-1)^{\ell(N-1)-(N+1)/2}}{\Gamma((N-1)(\ell-1/2))}
\end{equation}
This result agrees with the answer obtained by Davydychev and Smirnov 
\cite{DS} and reduces to the well-known four-dimensional behaviour found 
by Hagedorn and Almgren {\cite{HAK}, when we set $\ell=2$, namely
$$\rho_N(p) \sim Q^{(3N-5)/2},$$
where $Q = \sqrt{p^2}-\sigma$ is the energy release. However (6) supplies
the answer in any dimension with the appropriate coefficient.

\section*{III. Exact results}
We now consider phase space in general and derive some precise results for 
any $\ell,N$ as they arise from recurrence relations between phase space 
expressions for smaller $N$ and not necessarily around threshold. Begin with
the definition of the $N$-body phase space integral,
\begin{equation}
\rho_N(p)\equiv \rho_{p\rightarrow 1+2+\cdots N}\equiv\prod_i\left(
 \int\bar{d}^{2\ell}\!p_i\,\theta(p_i)\bar{\delta}(p_i^2-m_i^2)\right) 
 \bar{\delta}^{2\ell}(p-\sum_{i=1}^N p_i).
\end{equation}
The measures $d^{2\ell-1}\vec{p}\!=\!|\vec{p}|^{2\ell-2}d|\vec{p}|.
  (\sin\,\theta)^{2\ell-3}d\theta.2\pi^{\ell-1}/\Gamma(\ell\!-\!1)\!
=\!|\vec{p}|^{2\ell-2}d|\vec{p}|.2\pi^{\ell-1/2}/\Gamma(\ell\!-\!1/2)$, come
in useful if one were able to integrate over angles. Thus the two-body
result is readily evaluated in this way to be
$$\rho_2(p)=\frac{\pi(4\pi)^{1/2-\ell}q^{2\ell-3}}{\Gamma(\ell-1/2)
       \sqrt{p^2}},$$
where $q$ is the centre of mass spatial momentum, so that
$\sqrt{q^2+m_1^2}+\sqrt{q^2+m_2^2}=\sqrt{p^2}$. Tidying up, the result can 
be expressed covariantly as
\begin{equation}
 \rho_{p\rightarrow 1+2}=\frac{\pi^{1-\ell}\Gamma(\ell-1)\lambda^{2\ell-3}}
                              {2^{2\ell-1}(p^2)^{\ell-1}\Gamma(2\ell-2)};
 \quad \lambda\equiv\sqrt{p^4+m_1^4+m_2^4-2m_1^2m_2^2-2p^2m_1^2-2p^2m_2^2}.
\end{equation}

The three-body phase space can also be evaluated by brute force methods
and reduced to a triangular integral over three Mandelstam variables:
$$\rho_{p\rightarrow 1+2+3}=\frac{2\pi (p^2)^{1-\ell}}
       {(4\pi)^{2\ell}\Gamma(2\ell-2)}\int\!\!\int\!\!\int\!\! ds\,dt\,du\,
   \delta(s+t+u-m_1^2-m_2^2-m_3^3-p^2)[\Phi(s,t,u)]^{\ell-2}\theta(\Phi), $$
where $\Phi(s,t,u)$ is the Kibble \cite{K} cubic (simply a Gram determinant),
$$\Phi(s,t,u)\equiv stu-s(m_2^2m_3^2+p^2m_1^2)-t(m_3^2m_1^2+p^2m_2^2)
   -u(m_1^2m_2^2+p^2m_3^2)$$
$$\qquad\qquad\qquad\qquad +2(m_1^2m_2^2m_3^2+p^2m_1^2m_2^2+p^2m_2^2m_3^2
 +p^2m_3^2m_1^2)=-p^2\lambda^2(|\vec{p}_1|^2,|\vec{p}_2|^2,|\vec{p}_3|^2).$$
By changing variables to $s$ and $t-u$, one integration may be performed and
the problem reduced to the single integral \cite{BBBB},
\begin{equation}
\rho_{p\rightarrow 1+2+3}=\frac{(32\pi)^{2-2\ell}}
                         {(\Gamma(\ell-1/2))^2(p^2)^{\ell-1}}
\int_{(m_1+m_2)^2}^{(\sqrt{p^2}-m_3)^2} s^{1-\ell}{\cal D}^{\ell-3/2}\,ds,
\end{equation}
where ${\cal D}=[s-(m_3+\sqrt{p^2})^2][s-(m_3-\sqrt{p^2})^2][s-(m_1+m_2)^2]
[s-(m_1-m_2)^2]$.
When we are in four dimensions ($\ell=2$) the integral is nothing but the
area within the Dalitz plot but its evaluation for arbitrary masses is
easier said than done, because (9) is an elliptic function in general!
However in three dimensions the integration over the region becomes possible
because it can be converted into
$$\rho_3(p)\rightarrow\frac{1}{32\pi\sqrt{p^2}}
 \int_{(m_1+m_2)^2}^{(\sqrt{p^2}-m_3)^2}\frac{ds}{\sqrt{s}}
 =\frac{1}{16\pi}\left(1-\frac{m_1+m_2+m_3}{\sqrt{p^2}}\right) .$$
In fact the three-body phase space integration is tractable for all odd $D$ 
dimensions, because one remains with a polynomial in half-integral powers of
$s$ which is readily handled \cite{R}. Two other cases are amenable to an 
exact treatment in terms of `elementary' functions for any $\ell$ and $N=3$, 
namely (i) two masses are set to zero, or (ii) one mass vanishes and the two
other masses are equal. In case (i) put $m_2=m_3=0, m_1=m~$ so $~{\cal D}
\rightarrow (s-p^2)^2(s-m^2)^2$ whereupon the 3-body phase space reads
$$\rho_3(p)\rightarrow \frac{\pi^{1-2\ell}(p^2/m^2-1)^{4\ell-5} 
        [\Gamma(\ell-1)]^2}{2^{4\ell-1}(m^2)^{3-2\ell}\Gamma(4\ell-4)}
        F(3\ell-3,2\ell-2;4\ell-4;1-\frac{p^2}{m^2})\,\theta(p^2-m^2),$$
while in case (ii) put $m_3=0, m_1=m_2=m~$ so that $~{\cal D}\rightarrow
s(s-p^2)^2(s-4m^2)$ and
$$\rho_3(p)\rightarrow\frac{\pi^{3/2-2\ell}\Gamma(\ell-1)\,\,
    (p^2-4m^2)^{3\ell-7/2}}{2^{6\ell-3}\Gamma(3\ell-5/2)\,\,(p^2)^{\ell-1}m}
      F(\frac{1}{2},\ell-\frac{1}{2},3\ell-\frac{5}{2};1-\frac{p^2}{4m^2})
      \,\theta(p^2-4m^2).$$
Of course we can also proceed to the limit $m\rightarrow 0$ in either case.
Any other set of masses produces `elliptic' results \cite{HAK,BBBB} for 
even $D$ or integral $\ell$.

A more systematic way of arriving at Lorentz invariant answers, without 
resorting to integrations over spatial momenta in the standard approach and 
interpreting them covariantly, is to apply Almgren's method. In that method
one partitions the set of outgoing particles into subsets (call them $A, B$ 
etc.) and obtains $\rho$ via mass integration convolutions;
for instance the three-body result (9) can be construed as 
$$\rho_{p\rightarrow 1+2+3} = \int \rho_{p\rightarrow A+3}\,\,
                             \rho_{A\rightarrow 1+2}\,\bar{d}m_A^2,$$
where $s$ is interpreted as the intermediate $m_A^2$. More generally, one
may evaluate the $N$-body phase space through the double integral
\begin{equation}
\rho_{p\rightarrow 1+2+\cdots+j+(j+1)\cdots+N}=\int \rho_{p\rightarrow A+B}
 \,\rho_{A\rightarrow 1+2+\cdots j}\,\rho_{B\rightarrow(j+1)+\cdots+N}\,   
 \bar{d}m_A^2\,\bar{d}m_B^2.
\end{equation}
Although this is quite a satisfactory numerical way of calculating $\rho$,
it does not shed a great deal of analytical light on the nature of the
problem; but it does serve as a nice check of analytical results obtained in
a different manner, which we will now exhibit.

\section*{IV. Coordinate space method}

The sunset diagram having all $\nu$=1 is nothing but the $2\ell$-dimensional
Fourier transform of the product of $N$ causal functions:
\begin{equation}
 I_N(p) = -i\int d^{2\ell}\!x\,\exp(ip.x)\prod_{i=1}^N[i\Delta_c(x|m_i)],
\end{equation}
and the phase space integral is simply given by 
$$\rho_{p\rightarrow 1+2+\cdots+N}=2 \Im I_N(p),$$
which is nonvanishing for $p^2\geq (m_1+m_2+\cdots +m_N)^2$. Of course (11) 
is easier stated than done except in the simplest cases (like $N=2$ or 3)
because the causal function is a Bessel function in general \cite{AD},
$$i\Delta_c(x|m)= \frac{1}{(2\pi)^\ell}\left(\frac{m}{r}\right)^{\ell-1}
                  K_{\ell-1}(mr);\qquad r\equiv \sqrt{-x^2+i\epsilon}.$$
In the massless limit, when $m\rightarrow 0$, the sunset diagram reduces
to a ``superpropagator'' with integer index since
$$i\Delta_c(x|0)\equiv iD_c(x) = \Gamma(\ell-1)/4\pi^\ell r^{2\ell-2}.$$
However, as noted and indeed emphasized by Berends {\em et al} \cite{BBBS}
and by Groote {\em at al} \cite{GKP}, one can make considerable progress 
using coordinate space methods in the odd-dimensional massive cases, 
because the Bessel function reduces to an exponential times a polynomial.

Let us begin systematically by considering the most trivial case {\em when 
all intermediate masses vanish}. Carrying out the angular integration in (11),
the sunset integral reduces to
\begin{equation}
 I_N(p)\rightarrow -(2\pi)^\ell q^{1-\ell}\int_0^\infty\!dr\,\,\,r^\ell
           J_{\ell-1}(qr)\left[\frac{\Gamma(\ell-1)}{4\pi^\ell r^{2\ell-2}}
           \right]^N; \qquad q^2\equiv -p^2,
\end{equation}
which can be evaluated straightforwardly \cite{GRMOS}. Thus one finds that
\begin{equation}
 \lim_{m_i\rightarrow 0}I_N(p)= -(4\pi)^{\ell(1-N)}
           \frac{[\Gamma(\ell-1)]^N \Gamma(\ell+N-N\ell)}{\Gamma(N\ell-N)}
           (-p^2-i\epsilon)^{N\ell-\ell-N},
\end{equation}
so the massless phase space integral reduces, in $2\ell$ dimensions,
to\footnote{Here and afterwards, we use the fact that 
$\Im\Gamma(a)(-k^2+i\epsilon)^a=-\pi|k^2|^a\theta(k^2)/\Gamma(1-a)$ for real
$a$.}
\begin{equation}
 \lim_{m_i\rightarrow 0}\rho_N(p)=2\Im I_N(p)=\frac{(4\pi)^{1+\ell-N\ell}
     [\Gamma(\ell-1)]^N}{2\Gamma(N\ell-N)\Gamma((N-1)(\ell-1))}
     (p^2)^{N\ell-\ell-N}\theta(p^2).
\end{equation}
Both (13) and (14) are exact too. One may readily check that the cases 
$N=1,2,3$ produce the correct answers, by taking appropriate limits of 
earliers results. Also it is very instructive to check that the Almgren 
recurrence formulae, such as (10), are properly obeyed, not only in their 
momentum dependence, but in their multiplicative coefficients.

Now it turns out that the integral of two Bessel functions with an 
exponential can also be handled \cite{GRMOS}. Therefore one can improve on 
above and calculate analytically the sunset integral for one particle 
massive ($m$) and the remaining $N$-1 particles massless. In those 
circumstances one is led to \cite{BD}
\begin{eqnarray}
 I_N(p)&=&\!\!-\left(\frac{q}{m}\right)^{1-\ell}\!\!\int_0^\infty\!dr\,\,
        r^{2(N-1)(1-\ell)+1}J_{\ell-1}(qr)K_{\ell-1}(mr).
        \left(\frac{\Gamma(\ell-1)}{4\pi^\ell}\right)^{N-1}\nonumber\\
&=&\!\!-\frac{[\Gamma(\ell\!-\!1)]^{N\!-\!1}\Gamma(\ell\!+\!N\!-\!N\ell)
         \Gamma(N\!-\!N\ell\!+\!2\ell\!-\!1)}
         {2^{2\ell(N-1)}\pi^{\ell(N-1)}(m^2)^{N+\ell-N\ell}\Gamma(\ell)}
      F(\ell\!+\!N\!\!-\!\!N\ell,N\!\!-\!\!N\ell\!+\!2\ell\!\!-\!\!1;\ell;
     \!\frac{p^2}{m^2}\!).
\end{eqnarray}
Noting that in respect of the variable $z$ the hypergeometric function
$F(a,b;c;z)$ for real $a,b,c$ has a branch point at $z=1$ and that the 
discontinuity across the cut (which runs to +$\infty$) is
\begin{equation}
 \Im F(a,b;c;z) = -\frac{\pi\Gamma(c)(z-1)^{c-a-b}\theta(z-1)}
                {\Gamma(a)\Gamma(b)\Gamma(1-a-b+c)}F(c-a,c-b;c-a-b+1;1-z),
\end{equation}
we deduce the phase space result,
\begin{equation}
\rho_N(p)\!=\!\frac{\pi^{1\!-\!\ell(N\!-\!1)}[\Gamma(\ell\!-\!\!1)]^{N\!-\!1}
   (p^2\!\!-\!\!m^2)^{2N\ell\!-\!2\ell\!-\!2N\!+\!1}\theta(p^2\!\!-\!\!m^2)}
    {2^{2\ell(N-1)-1}(m^2)^{N\ell-N-\ell+1}\Gamma(2(N-1)(\ell-1))}
    F(\!N(\ell\!-\!1),(N\!-\!1)(\ell\!-\!1);2(\!N\!-\!1)(\ell\!-\!1);
      1\!-\!\frac{p^2}{m^2}\!)
\end{equation}
The pair of expressions (16) and (17) are precise as well. Contrast (17) 
with the expression found by Beneke and Smirnov \cite{BS} for a one-loop 
vertex diagram with two massive and one massless particle. By taking the 
limit $m\rightarrow 0$ of (16) and (17) and suitably manoeuvring the 
hypergeometric function\footnote{e.g. using the relation $F(a,b;c;z)=
(1-z)^{-b}F(c-a,b;c;z/(z-1))$.} one can show that they collapse into (13) 
and (14). Additionally we may deduce the particular cases $N=2,3$ 
previously by direct substitution; for instance, in four dimensions 
($\ell=2$), one gets
$$\rho_N(p) = \frac{2\pi(p^2-m^2)^{2N-3}}
                 {(16\pi^2)^{N-1}(m^2)^{N-1}\Gamma(2N-2)}
              F(N,N-1;2(N-1);1-\frac{p^2}{m^2})\theta(p^2-m^2),$$
so the three-body phase space with a single massive particle ($p^2\geq m^2$)
brings in a logarithmic function:
$$\rho_3(p)=\frac{(p^2-m^2)^3}{768\pi^3\,m^4}F(3,2;4;1-\frac{p^2}{m^2})
           = \frac{m^2}{256\pi^3 p^2}\left[(\frac{p^4}{m^4}-1)-
                    2\frac{p^2}{m^2}\ln(\frac{p^2}{m^2})\right].$$

\section*{V. Hypergeometric Expansions}
As soon as we have at least two massive particles, we come across integrals 
involving products of three Bessel functions with different arguments and
a power of $r$. This is not given in the standard texts\footnote{
The formula $2\!\int_0^\infty\!\!r^{\nu+1}K_\mu(ar)K_\mu(br)J_\nu(cr)
 dr\!=\!\!\sqrt{\frac{\pi}{2}}\left(\frac{c}{ab}\right)^{\nu+1}\!\!
\frac{\Gamma(\nu+\mu+1)\Gamma(\nu-\mu+1)}{\Gamma(\nu+3/2)(1+z)^{\nu+1/2}}
 F(1/2\!\!-\!\!\mu,1/2\!\!+\!\!\mu;3/2\!\!+\!\!\nu;(1\!\!-\!\!z)/2)$, where 
$z\equiv(a^2\!+\!b^2\!+\!c^2)/2ab,$ is given in \cite{PBM}.}, 
though it is surely some generalization of a hypergeometric function for we
definitely know how to evaluate the $N=2$ case directly in momentum space; 
but for larger $N$ it would seem that the coordinate space method stalls. 
Only when $p=0$, which corresponds to a vacuum `watermelon diagram', can 
the integrations sometimes be carried out \cite{GKP2}. However, we will 
now present a method which is still suitable for handling the general 
problem and which works wonderfully well in odd-dimensional spaces. 
It relies on the observation that the modified Bessel function possesses 
an asymptotic representation ($\mu\equiv 4\nu^2$),
$$K_\nu(z)\sim\sqrt{\frac{\pi}{2z}}{\rm e}^{-z}\left[1+\frac{\mu-1}{8z}+
          \frac{(\mu-1)(\mu-9)}{2!(8z)^2}+\frac{(\mu-1)(\mu-9)(\mu-25)}
               {3!(8z)^3}+\cdots\right] $$
that {\em terminates for $\nu$ or $\ell$ half-integral}, corresponding to 
$D$ odd. Even for integral $\ell$ or even dimensions, we shall see that it 
provides a very nice threshold expansion in terms of hypergeometric 
functions \cite{GP}. To show how this works, assume first that all the 
intermediate particles are massive. Since
\begin{equation}
 i\Delta_c(x|m)=\frac{1}{(2\pi)^\ell}\left(\frac{m}{r}\right)^{\ell-1}
                \sqrt{\frac{\pi}{2mr}}{\rm e}^{-mr}\sum_{j=0}^\infty
                \frac{\Gamma(\ell+j-1/2)}{j!\Gamma(\ell-j-1/2)}(2mr)^{-j},
\end{equation}
the product of causal Green functions produces the series
$$\prod_{j=1}^Ni\Delta_c(x|m_j)=\left(\frac{\pi^{1/2-\ell}}{2^{\ell+1/2}
                r^{\ell-1/2}}\right)^N\prod_{j=1}^N m_j^{\ell-3/2}.
                {\rm e}^{-r\sum_j m_j}\left[1+\frac{(2\ell-1)(2\ell-3)}{8r}
                    \sum_{j=1}^N\frac{1}{m_j}+\cdots \right].$$
The leading (and in 3-D the only) term of the rhs above yields the leading 
behaviour of the sunset diagram,
\begin{eqnarray}
 I_N^0(p)&=&\!\!-\frac{(2\pi)^{\ell-N\ell+N/2}}{q^{\ell-1}\varpi^{3/2-\ell}}
          \int_0^\infty dr\,\,r^{\ell-N\ell+N/2}{\rm e}^{-r\sigma}
          J_{\ell-1}(qr)\nonumber\\
          &=&\!\!-\frac{2^{1-N\ell-N/2}\pi^{\ell\!-\!N\ell\!+\!N/2}
              \Gamma((\!2\!-\!N\!)\ell\!+\!N/2\!)}
            {\Gamma(\ell)\sigma^{(2-N)\ell+N/2}\varpi^{3/2-\ell}}
  F(\!\frac{2\ell\!-\!N\ell\!+\!N/2}{2},\frac{2\ell\!-\!N\ell\!+\!N/2\!+\!1}
            {2};\ell;\!\frac{p^2}{\sigma^2}\!),
\end{eqnarray} 
where $\varpi\equiv \prod_im_i,\,\,\sigma\equiv\sum_im_i$. The imaginary
part then produces the leading contribution to the phase space integral,
\begin{eqnarray}
 \rho_N^0(p)&=&\frac{2^{2\ell(1-N)+1}\pi^{\ell(1-N)+(N+1)/2}
                  (p^2\!-\!\sigma^2)^{\ell(N-1)-(N+1)/2}}{\varpi^{3/2-\ell}
                  \sigma^{N\ell-1-N/2}\Gamma(\ell(N-1)-(N+1)/2)}
                  \theta(p^2-\sigma^2)\times\nonumber\\
  & &  F(\frac{N\ell-N/2}{2},\frac{N\ell-1-N/2}{2};\ell(N-1)-\frac{N-1}{2};
           1-\frac{p^2}{\sigma^2}).
\end{eqnarray}
It should be noted that expressions (19) and (20) represent the {\em
complete} results for 3-D, when they reduce respectively to
$$I_N(p)\rightarrow-\frac{\pi^{3/2-N}\Gamma(3-N)}
                         {2^{2N-2}\sigma^{3-N}\sqrt{\pi}}
         F(\frac{3-N}{2},\frac{4-N}{2};\frac{3}{2};\frac{p^2}{\sigma^2})$$
and
$$\rho_N(p)\rightarrow\frac{\pi^{2-N}(p^2-\sigma^2)^{N-2}}
                       {2^{3N-4}\sigma^{N-1}\Gamma(N-1)}
                F(\frac{N}{2},\frac{N-1}{2};N-1;1-\frac{p^2}{\sigma^2})
 = \frac{\pi(\sqrt{p^2}-\sigma)^{N-2}\theta(p^2-\sigma^2)}
        {(4\pi)^{N-1}\Gamma(N-1)\sqrt{p^2}}.$$

Nonleading behaviours (for values of $\ell \neq 3/2$) of $I_N$ and $\rho_N$
can be found from expansion (18); for instance to the next order we 
encounter the terms,
\begin{eqnarray}
 I_N^1(p)&=&\frac{(2\ell-1)(2\ell-3)\pi^{\ell((1-N)+N/2}
                \Gamma(2\ell-1-N\ell+N/2)}{2^{2+N\ell+N/2}\varpi^{3/2-\ell}
                \sigma^{\ell(2-N)+N/2-1}\Gamma(\ell)\mu}\times\nonumber\\
 & &\qquad F(\frac{2\ell-N\ell+N/2-1}{2},\frac{2\ell-N\ell+N/2}{2};\ell;
                      \frac{p^2}{\sigma^2}),
\end{eqnarray}
\begin{eqnarray}
 \rho_N^1(p)&=&\frac{(2\ell-1)(2\ell-3)\pi^{\ell(1-N)+(N+1)/2}
                     (p^2-\sigma^2)^{\ell(N-1)+(1-N)/2}}{2^{2\ell(N-1)+3}
        \varpi^{3/2-\ell}\sigma^{N(\ell-1/2)}\mu\Gamma(\ell(N-1)+(3-N)/2)}
        \theta(p^2-\sigma^2)\times\nonumber\\
& &\qquad F(\frac{N\ell+1-N/2}{2},\frac{N\ell-N/2}{2};
        \ell(N-1)+\frac{3-N}{2}; 1-\frac{p^2}{\sigma^2}),
\end{eqnarray}
where $1/\mu\equiv \sum_j 1/m_j$; and so on for higher $I_N^k,\rho_N^k$. The
method is entirely systematic and one may proceed to as high an order $k$ as
needed. Observe that the resulting series is a sort of threshold expansion 
because the powers of $(p^2-\sigma^2)$ which multiply $F(a,b;c;
1-p^2/\sigma^2)$ keep on increasing as we raise $k$ while the denominators 
are associated with sums of the type $\sum_j(1/m_j)^k$.

However these expansions are deficient in one respect: it is very tricky to
consider the limit as one or several masses vanishes, because the 
approximation (18) is fairly useless for massless propagators; in that case 
we should be using directly $iD_c(x)\propto r^{2-2\ell}$, rather than the
asymptotic expansion (18). 
Finally then we will consider the case where $M$ of the $N$ particles are 
massive while the remaining $N-M$ are massless; this covers essentially all 
cases of interest. The sunset integral is given {\em ab initio} by
\begin{eqnarray}
I_N(p)&=&-\frac{(2\pi)^\ell}{q^{\ell-1}}\int_0^\infty dr\,r^\ell J_{\ell-1}
    (qr)\left(\!\frac{\Gamma(\ell-1)}{4\pi^\ell r^{2\ell-2}}\!\right)^{N-M}
        \!\!\left(\frac{(\pi r)^{1/2-\ell}}{2^{\ell+1/2}}\right)^M
         \!\!\!\!{\rm e}^{-r\sigma}\varpi^{\ell-3/2}\times\nonumber\\
&&\qquad\qquad\qquad\left[1+\frac{(2\ell-1)(2\ell-3)}{8r\mu} +\cdots\right],
\end{eqnarray}
where the symbols now refer simply to the massive intermediate particles; 
therefore $\varpi\equiv\prod_{j=1}^M m_j$,  $\sigma\equiv\sum_{j=1}^M m_j,\,
1/\mu\equiv\sum_{j=1}^M (1/m_j)$. The integrals in (23) are readily performed
and the leading term of the full answer is
\begin{eqnarray}
 I_N^0(p)&\!\!=&\!-\frac{2^{2(M-N)+1-M\ell-M/2}\pi^{\ell(1-N)+M/2}
      [\Gamma(\ell\!-\!1)]^{N-M}\Gamma(\ell(2-M)+(M-N)(2\ell-2)+M/2)}
      {\varpi^{3/2-\ell}\sigma^{\ell(2\!-\!M)+(M\!-\!N)(2\ell\!-\!2)+M/2}
      \Gamma(\ell)} \nonumber\\
 & & \times F(\frac{\ell(2\!-\!M)\!+\!M/2}{2}+(M\!-\!N)(\ell\!-\!1),
     \frac{\ell(2\!-\!M)\!+\!1\!+\!M/2}{2}\!+\!(M\!-\!N)(\ell\!-\!1);
        \ell;\frac{p^2}{\sigma^2}).
\end{eqnarray}
Taking its discontinuity, the leading phase space behaviour is
 ($c\equiv M(3/2-\ell)+2N(\ell-1)-\ell+1/2$ and $p^2\geq\sigma^2$ below),
\begin{eqnarray}
\rho_N^0(p)&=&\frac{2^{2\ell(1-N)+1}\pi^{\ell(1-N)+(M+1)/2}
  [\Gamma(\ell-1)]^{N-M}(p^2-\sigma^2)^{M(3/2-\ell)+2N(\ell-1)-(\ell+1/2)}}
  {\varpi^{3/2-\ell}\sigma^{M(7/2-3\ell)+2(M+N)(\ell-1)-1}
   \Gamma(c)}\times\nonumber \\
& & F((\ell\!-\!1)(N\!-\!M)\!+\!\frac{M(2\ell-1)}{4},
       (\ell\!-\!1)(N\!-\!M)\!+\!\frac{M(2\ell-1)}{4}\!-\!\frac{1}{2};
       c;1-\frac{p^2}{\sigma^2}).
\end{eqnarray}
All the previous cases fall out of (24) and (25) by making the relevant 
substitutions for $M$ and $\ell$. Of course one may also derive nonleading
terms $I_N^k,\rho_N^k$ in exactly the same way as before and they correspond
to higher order threshold corrections. Again, these corrections entrain
higher powers of $(p^2-\sigma^2)$ and terminate for half-integral $\ell$.

\section*{VI. Conclusions}
Before we can comprehend the statistical effects of phase space on brane 
physics, it is vital to understand phase space in flat $D$-dimensional 
spacetime for any number of particles, with arbitrary masses. This paper 
has been devoted to that subject and we have arrived at results for 
$\rho_N(p)$ and $I_N(p)$ that have culminated in formulae (14), (23), (24) 
and (25). These comprise the high-energy and low-energy characteristics and
at any energy in-between. We believe that these analytical expressions are 
as compact as one can make them and will turn out to be practically useful. 
Otherwise one will be obliged to resort to numerical methods.

\section*{Acknowledgements}We are grateful to Dr AI Davydychev for supplying
us with a comprehensive list of references in this vast topic, for pointing 
us in the right direction and for valuable comments on our work. We also thank
the Australian Research Council for financial support under grant number 
A00000780.

\newpage
\section*{Appendix. At the threshold of a sunset}

In this appendix we show explicitly how the threshold expansion can be
carried out in momentum space via Feynman parameters. We illustrate the case
$N=2$ for any dimension $2\ell$ to keep the algebra simple, when the sunset
integral is just
$$I_2(p) = \frac{\Gamma(2-\ell)}{(-4\pi)^\ell}\int_0^1\!d\alpha_1
           \int_0^1\!d\alpha_2\,\frac{\delta(\alpha_1+\alpha_2-1)}
            {[p^2\alpha_1\alpha_2-m_1^2\alpha_2-m_2^2\alpha_1]^{2-\ell}}.$$
Never mind the fact that $I_2$ can be expressed as a linear combination of 
two hypergeometric functions ($a=\ell-1,b=2-\ell,c=\ell$) with arguments
$\frac{1}{2}[(m_1^2-m_2^2\pm p^2)/\lambda+1]$ or that it can be accorded
an exact geometrical interpretation \cite{DD}; our purpose here is to show 
how one may arrive at a systematic threshold expansion, by expanding about 
the minimum of the denominator, regarded as a function of $\alpha$. Thus
change variables to
$$\alpha_i = \frac{m_i}{m_1+m_2}+\Delta\rho_i,\quad 
  p^2\equiv(m_1+m_2)^2[1-\Delta^2].$$
The delta-function constraint on $\alpha$ corresponds to $\rho_1 =-\rho_2
\equiv \rho$, whereupon the sunset integral reduces to
$$I_2(p)=\frac{\Gamma(2-\ell)}{(4\pi)^\ell}\Delta^{2\ell-3}
         \int_{-\frac{m_1}{(m_1+m_2)\Delta}}^{\frac{m_2}{(m_1+m_2)\Delta}}
         \frac{d\rho}{[m_1m_2+\rho^2(m_1+m_2)^2(1-\Delta^2)-
                       \Delta\rho(m_1^2-m_2^2)]^{2-\ell}}.$$
The leading behaviour is found by setting $\Delta=0$ in the denominator
of the integrand and extending the limits to $\pm\infty$. Since
$$\int_0^\infty\!\!\! \frac{d\rho}{[m_1m_2\!+\!\rho^2(m_1+m_2)^2]^{2-\ell}}
  \!= \frac{(m_1m_2)^{\ell-3/2}}{2(m_1+m_2)}\int_0^\infty 
            \!\! \frac{dv\,v^{-1/2}}{(1\!+\!v)^{2-\ell}}\!=\!
  \frac{(m_1m_2)^{\ell-3/2}}{2(m_1+m_2)}
  \frac{\sqrt{\pi}\Gamma(3/2-\ell)}{\Gamma(2-\ell)},$$
it follows that
$$I_2^0(p)=\frac{(m_1m_2\Delta^2)^{\ell-3/2}\sqrt{\pi}}
         {(4\pi)^\ell(m_1+m_2)}\Gamma(3/2-\ell).$$

However we are keen to obtain all the nonleading terms and this can be 
accomplished by rewriting the sunset integral as
$$I_2(p)=\frac{\Gamma(2-\ell)(m_1m_2\Delta^2)^{\ell-3/2}}
       {(4\pi)^\ell\sqrt{p^2}}\int_{u_-}^{u_+}du\,[1+u^2-u\eta]^{\ell-2},$$
where the new integration variable is $u=\rho\sqrt{p^2/m_1m_2}$, the limits
are $u_+=\frac{m_2}{(m_1+m_2)\Delta}\sqrt{\frac{p^2}{m_1m_2}}$ and 
$u_-=-\frac{m_1}{(m_1+m_2)\Delta}\sqrt{\frac{p^2}{m_1m_2}}$, and the new
expansion variable is $\eta\equiv\frac{\Delta(m_1^2-m_2^2)}
{\sqrt{p^2m_1m_2}}$. The series in $\eta$ reads
$$I_2(p)=\frac{(m_1m_2\Delta^2)^{\ell-3/2}}{(4\pi)^\ell\sqrt{p^2}}
         \sum_{n=0}^{\infty}\frac{\Gamma(2+n-\ell)}{n!}\eta^nU_n;\qquad
        U_n\equiv \int_{u_-}^{u_+}du\,\,\,(1+u^2)^{\ell-2-n}u^n.$$
In evaluating the coefficients $U_n$, we make use of the fact that
the limits are large and we should distinguish between the cases of $n$ 
even and $n$ odd. For $n$ even, we split the integral into
\begin{eqnarray*}
U_n&=&\left(2\int_0^\infty-\int_{|u_-|}^\infty-\int_{|u_+|}^\infty\right)
       du \,\, u^n(1+u^2)^{\ell-2-n}\\
 &=& \frac{\Gamma((n+1)/2)\Gamma((3+n)/2-\ell)}{\Gamma(2+n-\ell)}-
    \left(\int_{|u_-|}^\infty+\int_{|u_+|}^\infty\right)du\,
              u^{2\ell-4-n}(1+1/u^2)^{\ell-2-n},
\end{eqnarray*}
where we may expand the latter two integrals in powers of $1/u^2$ --- which
provides a further sub-expansion in powers of $\Delta$. Thus we get
$$\Gamma(2+n-\ell)U_n=\Gamma((n+1/2)\Gamma((3+n)/2-\ell)+
  \sum_{k=0}^\infty\frac{(-)^k\Gamma(2+n+k-\ell)}{k!(2\ell-3-n-2k)}\times$$
$$\qquad\qquad\qquad\qquad\qquad\qquad\qquad
  \left[|u_+|^{2\ell-3-n-2k}+|u_-|^{2\ell-3-n-2k}\right].$$
On the other hand, the case of odd $n$ can be treated directly by expanding 
the integrand below in powers of $1/v$:
\begin{eqnarray*}U_n&=&\frac{1}{2}\int_{u_-^2}^{u_+^2}dv\,\,v^{\ell-(5+n)/2}
                   (1+1/v)^{\ell-2-n} \qquad {\rm so}\\
 \Gamma(2+n-\ell)U_n&=& \sum_{k=0}^\infty\frac{(-)^k\Gamma(2+n+k-\ell)}
  {k!(2\ell-3-n-2k)}\left[|u_+|^{2\ell-3-n-2k}-|u_-|^{2\ell-3-n-2k}\right].
\end{eqnarray*}
Again one encounters a sub-series in $\Delta$. (The two series make up the
`h-h' and `p-p' contributions of ref \cite{DS} at one stroke.) In short we 
see that the sunset integral, and of necessity its imaginary part, can be
systematically expanded as a series in $\Delta$ which is proportional to
the $Q$-value of the reaction by appropriately handling 
the Feynman parametric representation. It is possible to treat the
sunset diagram with $N$ intermediate particles in a similar way, but
the method becomes rather unwieldy, which is why we turned to coordinate 
space methods in sections IV, V.

\end{document}